\documentclass[12pt,preprint]{aastex}
\usepackage{epsfig}

\shorttitle{A near-infrared counterpart to 1E~2259+58.6?}
\shortauthors{Hulleman et~al.}

\begin{document}

\title{A possible faint near-infrared counterpart to the AXP 1E~2259+58.6}

\author{F.~Hulleman,\altaffilmark{1} A.~F.~Tennant,\altaffilmark{2}
M.~H.~van Kerkwijk,\altaffilmark{1} S.~R.~Kulkarni,\altaffilmark{3}
C.~Kouveliotou,\altaffilmark{4,2} and S.~K.~Patel\altaffilmark{4}}
\altaffiltext{1}{Utrecht University, P. O. Box 80000,
3508 TA~~Utrecht, the Netherlands}
\altaffiltext{2}{NASA Marshall Space Flight Center, Huntsville, AL 35812}
\altaffiltext{3}{Palomar Observatory, California Institute of
Technology 105-24, Pasadena, CA 91125} 
\altaffiltext{4}{Universities Space Research Association /NSSTC, SD-50,
Huntsville, AL 35805}

\begin{abstract}
We present near-infrared and optical observations of the field of the
Anomalous X-ray Pulsar 1E~2259+58.6 taken with the Keck telescope. We
derive a subarcsecond {\em Chandra} position and tie it to our optical
reference frame using other stars in the field. We find a very faint
source, $K_{\rm s} = 21.7\pm0.2$\,mag, with a position coincident with the
{\em Chandra} position.  We argue that this is the counterpart.  In
the J, I, and R bands, we derive ($2\sigma$) limits of 23.8, 25.6 and
26.4\,mag, respectively.  As with 4U~0142+61, for which a counterpart
has previously been found, our results are inconsistent with models
in which the source is powered by accretion from a disk, but may be
consistent with the magnetar model.
\end{abstract}


\section{Introduction}

1E~2259+58.6 is a member of a class of X-ray pulsars whose energy
source is not understood, called Anomalous X-ray Pulsars \citep[AXPs, see
][for a review]{israms02}. Their rotational energy loss is
insufficient to explain their X-ray luminosities and there is no evidence
for a companion star from which matter could be accreted.  

In part, the uncertainty about AXPs arises because of a lack
of counterparts at other wavelengths.  Therefore, we have initiated a
program of deep optical and near-infrared observations of those AXPs
which are not too highly absorbed.  This program has led to the
detection of the first optical counterpart of an AXP, namely the
counterpart to 4U~0142+61 \citep{hullvkk00}. 
Our earlier observations of 1E~2259+58.6 set stringent limits \citep[$R>25.7$,
$I>24.3$\,mag,][]{hullvkvk00}. 

In this {\it Letter}, we present in \S\ref{secobs} deeper optical
observations and new near-infrared observations of 1E~2259+58.6.  In
\S\ref{seccxo}, we derive an accurate position from {\em Chandra}
observations \citep[see the companion paper by][]{patekw+01}, which we
use in \S\ref{secid} to identify a possible near-infrared counterpart
to 1E~2259+58.6. 

\section{Observations}
\label{secobs}

\subsection{Optical observations}

Deep R and I band images of the field around 1E~2259+58.6 were taken
on the night of 2000 September 3, with the Low Resolution Imager
Spectrograph \citep[LRIS,][]{okecc+95} mounted on Keck I.  The night
was photometric and the seeing was $0\farcs6$ ($0\farcs9$) in R (I).
Six (twelve) images were averaged for a total integration time of 1
hour in R (I). Less deep, wider field of view, B, V, R, and I images
were taken at the 60-inch telescope on Palomar mountain on the night
of 2000 November 18 and were used for the astrometry (\S\ref{seccxo}).

The reduction was done using the ESO-MIDAS software package, in the
same way as described by \citet{hullvkvk00,hullvkk00}.
The stacked Keck I band image is shown in Fig.~\ref{fig1}, the Keck R band
image is not shown. It is very similar to the image shown in
\citet{hullvkvk00}. 

We measured instrumental magnitudes on the Keck images using
point-spread-function (PSF) fitting as implemented in the 
DAOPHOT package, and calibrated these using the calibration found
earlier \citep{hullvkvk00}.  We estimate single-trial 2$\sigma$
detection limits of 26.4 (25.6)\,mag in R (I) by adding 8.1 (7.6)\,mag
to the mean instrumental magnitude of objects that have a 
PSF fitting error of 0.3\,mag\footnote{The detailed simulations described in
\cite{hullvkvk00} convinced us that this procedure gives correct
estimates.}. 

\subsection{Near infrared photometry}

Near-infrared images in the J and K$_{\rm s}$ filters were obtained on
the nights of 1999 June 23 and 24, with the Near Infrared Camera
\citep[NIRC,][]{matts94}, mounted on the Keck I telescope. During both
nights observing conditions were photometric.  During the first night, the
seeing was mediocre, varying between $0\farcs6$ and
$1\farcs5$, but during the second night it was excellent, at
$0\farcs5$ in both K$_{\rm s}$ and~J.  Therefore, we only use the data taken
on the second night.  These consist of a total of 36 frames in each of
J and K$_{\rm s}$, with each $K_{\rm s}$ (J) frame composed of 5 (3)
coadded exposures of 12 (20) seconds.

The data were reduced using MIDAS. After subtracting dark frames, a
flatfield and a mosaic were created for each filter in the following
way. First, each individual frame was normalized to unity. Next, at
each pixel the variance over all the (normalized) frames was
computed. For a field which is not too crowded, the median of these
variances should be a good approximation of the expected variance of a
pixel which in each frame only contains sky.  Then, for each pixel we
used an F-test to determine if its variance was significantly higher
than the median variance, where we set the level of significance at
0.001. If the variance was not significantly higher, we simply took
the mean of all frames at that pixel. Otherwise, we discarded the
highest values at each pixel until the variance was consistent with
the median variance. From the values left, a new mean was calculated.

After division by this flatfield, we used the prescription of
\citet{liugr01} to correct each frame for a persistence effect in the
read-out amplifiers and subtracted the background, which was determined
from the frame itself. 

A few well detected stars were used to determine the spatial offsets
between individual frames. Next, at each sky position the median
and standard deviation were computed (excluding physical pixels which
were obviously bad). If at any sky position an individual frame
differed from the median by more than three times the standard
deviation, it was discarded and a new standard deviation was computed.
This was repeated, until the values of all remaining frames were
within three standard deviations of the median. The resulting mosaics,
shown in Fig.~\ref{fig1}, have the corresponding mean value at each
sky position.  

We used DAOPHOT to measure instrumental magnitudes on the
mosaics, and calibrated these using the infrared standard star SJ~9182
\citep{persmk+98}.  In Table~\ref{tab:phot}, we list positions and
magnitudes of selected objects (see Fig.~\ref{fig1}).  We also
estimated single-trial 2$\sigma$ detection limits of 23.8\,mag in J and
22.4\,mag in K$_{\rm s}$ in the way described above.

\subsection{Optical Astrometry}

We determined coordinates relative to the USNO-A2.0 catalog
\citep{monebc+00}.  For the 60-inch images, 1107 USNO-A2.0 stars
overlap.  We determine centroids for 942 objects that were not saturated
and appeared stellar, corrected these for instrumental distortion
(J.~Cohen, 1997, private comm.), and solved for zero-point position,
scale, and position angle.  After rejecting 193 outliers, the rms
residuals are $0\farcs30$ in both right ascension and declination.  
Therefore, the accuracy of the tie to the USNO-A2.0
system is very good, with a formal uncertainty of $0\farcs01$
(standard deviation of the mean). The real uncertainty may be limited by
uncertainties in the distortion correction; but we believe it should
still be better than $0\farcs03$.  

The optical Keck images are too deep and the infrared images cover too
small a field for a direct tie to the USNO-A2.0 catalog.  Instead, for
those, we used stars in common with a short R-band exposure, for which
\citet{hullvkvk00} had determined coordinates in the USNO-A2.0 system.
For all bands, the tie to the short R-band image is accurate to better
than $0\farcs02$, and the final tie to the USNO-A2.0 system should again be
better than 0\farcs03.  

\section{The X-ray position}
\label{seccxo}

1E~2259+58.6 was observed with {\em Chandra} on 2000 January 11,
using the Advanced CCD Imaging Spectrometer (ACIS).  The analysis of this
observation is described in the companion paper by
\citet{patekw+01}.  Here, we use only the data collected in Timed
Exposure mode to derive a position for 1E~2259+59.6. We used the data
that were ``reprocessed'' on 2000 October 31 through the standard
pipeline.  This reprocessing corrected minor errors in the aspect 
solution compared to the original processing.  Further, in a calibration
update made on 2001 July 26 the ACIS pixel size was changed from 0.492
arcsec to 0.49131 arcsec.  We applied this correction manually.

For 1E~2259+58.6, which is rather bright, the data suffer from
pile-up of photons in the image core, as a result of which the source
looks ring-shaped with a hole in the center.  We modeled the data
using a Gaussian multiplied by a hyperbolic tangent in radius, scaled
to approach zero at $r=0.0$.  The resulting best-fit centroids
correspond, using the {\em Chandra} aspect solution, to a J2000
position on the sky of $\alpha=23^{\rm h}01^{\rm m}08\fs26$,
$\delta=+58\arcdeg52\arcmin44\farcs9$.  The uncertainty is limited by
systematic effects, to a circle with $\sim\!0.7\arcsec$ radius
\citep{aldckc+00}. 

To improve this position, we tried to identify optical
counterparts to other sources in the field and use these to obtain a
boresight correction.  For this purpose, we use all seven other
sources detected by the source-finding algorithm in the central (S3)
chip (see Table~\ref{tab:cxosources}). Three of the sources (X1, X5,
and X6) have X-ray spectra with a soft component, characteristic of
stellar sources with little absorption. All of these have
counterparts on the Digitized Sky Survey (DSS) plates and positions in
the USNO-A2.0 catalog.  Object X1 is RP1 of
\citet{rhop97}; \cite{vdbev01} found a G7-K1 IV-III star with
$V=11.8\pm0.1$ coincident with RP1. The expected X-ray luminosity of
the star agrees with the one measured for RP1. The stars coinciding
with object X5 and X6 are relatively bright as well (USNOA2.0 blue and
red magnitudes of 17.5 and 16.6 (X5) and 17.7 and 15.3 (X6)); given
the small probability of finding stars this bright at the X-ray
position by chance, we conclude that these most likely are the optical
counterparts.   

Interestingly, no counterparts are seen on the DSS for any of the
other sources, which have harder X-ray spectra.  However, for X2,
which has a low absorbing column, we do find a faint blue star on our
Palomar images.  Given the unusual color, we believe this is the
counterpart.  Furthermore, for X7, which appears absorbed, we find a
faint extended object at the edge of our 1997 R and I-band LRIS
images.  Given the crowdedness of the field, we cannot be sure this is
the counterpart; if it is, X7 is most likely a highly reddened
extragalactic object.  Finally, source X4 also overlaps with the 1997
LRIS images, but is in a very confused region. 
 
In Table~\ref{tab:cxosources}, we list positions inferred from our
Palomar images, from the USNO-A2.0 catalog, as well as positions from
the DSS (both epochs) using astrometric calibrations relative to the
USNO-A2.0 catalogue.  None of the objects appear to have significant
proper motion.  We used X1 and X6 to establish the boresight
correction and X2 to verify our solution. We did not use X5 because it
is faint and relatively far off-axis in the ACIS image, and it is not
covered by our Palomar data.  

In the {\em Chandra} images, X1, X2, and X6 are off-axis and thus appear
elongated, which may bias the position derived by the source-finding
algorithm. For each X-ray source, we computed a model PSF that was
rotated by the spacecraft roll angle to match the image orientation
and fit this model to a two dimensional elliptical Gaussian.  The
offset of the center of this Gaussian from the origin tells us the
expected size of the image shift introduced by the Gaussian fit.  
We then fit a Gaussian with the same $\sigma_x$,
$\sigma_y$, and inclination to the actual data and subtract the offset
determined from the model PSF.  We also list the uncertainty in the
centroid location estimated from the fit to the actual data.

Armed with both optical and X-ray coordinates for these objects,
we adjusted the {\em Chandra} coordinate system using the constant offsets,
$\delta x$ and $\delta y$, that gave the best agreement with the
Palomar positions.  The resulting best J2000 position for 1E~2259+58.6 is
$\alpha=23^{\rm h}01^{\rm m}08\fs295$,
$\delta=+58\arcdeg52\arcmin44\farcs45$.  We stress that this position
is tied directly to our optical images, so the only source of
systematic uncertainty in the position is the uncertainty in the
offset, of $0\farcs19$ in each coordinate.
This dominates over the statistical uncertainty, and corresponds to a
99\% confidence error radius of~$0\farcs60$. The corrected {\em
Chandra} position is indicated in Fig.~\ref{fig1}. 

\section{Identification of a Counterpart}
\label{secid}

We searched for possible counterparts within the {\em Chandra} error
circle. We did not find any
counterpart in our R, I and J images and place upper limits of 26.4,
25.6 and 23.8\,mag, respectively.
The R and I limits are nearly a magnitude
better than our previously published values \citep{hullvkvk00}.
However, in the new near-infrared K$_{\rm s}$ image,
we find a $K_s=21.7\pm 0.2$\,mag object (star 1) within the
{\em Chandra} error circle (Fig.~\ref{fig1}).

Another object, star 2, while within
the ROSAT error circle \citep[see][]{hullvkvk00} is offset
from the Chandra position by $0\farcs85\pm0\farcs25$, where the error
includes the uncertainty in the optical position as well as the
systematic uncertainty in the X-ray to optical transformation. It is 
inconsistent with being the near-infrared counterpart of 1E~2259+58.6
at the 99.9\% confidence level.

It is possible that star 1 is an unrelated background object. The
density of such faint objects is 0.04\,arcsec$^{-2}$.
Hence, thanks to the exquisite astrometric precision of {\em Chandra}, 
the probability of chance coincidence is small, less than 3\%. 

Thus, most likely we have identified the near-infrared
counterpart to 1E~2259+58.6. \citet{patekw+01} report 
$N_H=9.3\pm0.3\times 10^{21}$ cm$^{-2}$ which, using $A_{\rm
V}=\frac{N_{\rm H}}{1.79\times10^{21}}$\,mag \citep{preds95}, corresponds to
the following extinction values: $A_{\rm R}=4.3$, $A_{\rm I}=3.1$,
$A_{\rm J}=1.4$, $A_{\rm K}=0.6$\,mag, which should be accurate to
at least a factor of two. Adopting these
values we show the optical through X-ray photometry of 1E~2259+58.6
in Fig.~\ref{fig2}.  Also shown are the blackbody
and power-law components of the best-fit model to the X-ray data
\citep{patekw+01}.  

As for 4U~0142+61, we find an extremely large X-ray to
optical/infrared flux ratio, which is hard to square with most models
proposed for AXP.  Models in which the source is powered by accretion
from a fossil disk \citep[e.g.,][]{chathn00} were already excluded by
our previous optical limits.  However, this was for a standard disk in
which optical emission is due to reprocessing of the X-ray flux.

\cite{hullvkk00} showed that for 4U~0142+61 a disk with an
inefficiently radiating inner part was excluded as well, but that a
truncated disk, such as might be present in a very tight binary, was
still possible.  For 1E~2259+58.6, we can also exclude this
possibility, since the K band flux and optical limits are inconsistent
with the predicted roughly Rayleigh-Jeans spectral-energy
distribution.  For the same reason, we can exclude the possibility
that 1E~2259+58.6 is a hot white dwarf.

The only model that is without problem, although mostly for lack of
predictions, is that of \cite{thomd96}, in which AXP are magnetars,
powered by the decay of a superstrong, $\sim\!10^{15}$\,G, magnetic
fields.  We hope that our work will stimulate further study into
these models. In this respect, we are encouraged by the work done on
the X-ray emission by e.g.\ \cite{pernhh+01} and \cite{ozelpk01} and the qualitative agreement between observations and
rough estimates of the expected optical flux made by C.~Thompson
(2001, private comm.). 

With two counterparts identified, observational progress should be
rapid.  It should be relatively easy to determine the broad-band
spectral energy distributions of both sources and identify other AXPs.  
More difficult, but more rewarding potentially, would be searches for
optical pulsations and polarisation.

\acknowledgements We thank David Kaplan for his help with the
near-infrared reductions, for undertaking the LRIS observations, and
for donating Palomar observing time.  We thank Bryan Jacoby for doing
the Palomar observations. The referee (G.L. Israel) is thanked for a careful
reading of the manuscript. Part of the observations reported here
were obtained at the W.~M.~Keck Observatory, which is operated by the
California Association for Research in Astronomy, a scientific
partnership among California Institute of Technology, the University
of California and NASA. It was made possible by the generous financial
support of the W.~M.~Keck Foundation.  Part of this work was done
while FH, MHvK and CK were visiting the Institute for Theoretical
Physics, which is supported by NSF grant PHY99-07949. FH thanks
the Netherlands Organization for Scientific Research and the
Leids Kerkhoven Bosscha Fund for grants that made this visit possible.  MHvK
acknowledges support from a fellowship of the Royal Netherlands
Academy of Science, SRK grants from NSF and NASA, and CK and SKP
grants NAG5-9591 and GO-1018X. 

\bibliographystyle{apj}
\bibliography{e2259}

\clearpage

\begin{table}[b]
\smallskip
\caption[]{
Positions and magnitudes of selected objects\label{tab:phot}}
\smallskip
\begin{tabular}{ccccc}
\tableline
\multicolumn{1}{c}{\lower3pt\hbox{$\stackrel{\displaystyle
{\rm ID}}{}$}}&
\multicolumn{1}{c}{\lower3pt\hbox{$\stackrel{\displaystyle
\alpha_{J2000}}{23\hbox{}01+}$}}& 
\multicolumn{1}{c}{\lower3pt\hbox{$\stackrel{\displaystyle
\delta_{J2000}}{+58\hbox{ }52+}$}}&
\multicolumn{1}{c}{\lower3pt\hbox{$\stackrel{\displaystyle
{\rm J}}{}$}}&
\multicolumn{1}{c}{\lower3pt\hbox{$\stackrel{\displaystyle
{\rm K_{\rm s}}}{}$}}\\
\tableline
A &08.027&37.90& $17.178\pm 0.012$& $16.349\pm 0.006$\\
B &07.323&48.67& $18.453\pm 0.012$& $17.642\pm 0.009$\\
B'&07.359&47.10& $16.910\pm 0.011$& $16.207\pm 0.008$\\
D &08.475&35.71& $17.871\pm 0.012$& $17.114\pm 0.007$\\
F &09.454&53.51& $16.609\pm 0.010$& $15.908\pm 0.009$\\
G &09.721&48.26& $16.449\pm 0.013$& $15.957\pm 0.010$\\
H &07.079&40.51& $19.763\pm 0.012$& $18.963\pm 0.015$\\
I &07.503&43.71& $20.021\pm 0.011$& $19.066\pm 0.014$\\
J &09.181&50.73& $20.038\pm 0.013$& $19.236\pm 0.018$\\
K &09.361&39.71& $19.497\pm 0.012$& $18.651\pm 0.015$\\
L &09.778&41.07& $20.165\pm 0.017$& $19.022\pm 0.013$\\
L'&09.691&41.38& $21.110\pm 0.034$& $19.747\pm 0.028$\\
M &07.577&38.41& $20.868\pm 0.017$& $19.648\pm 0.027$\\
N &08.844&39.90& $19.950\pm 0.012$& $19.037\pm 0.014$\\
1 &08.312&44.53& \nodata          & $21.670\pm 0.193$\\
2 &08.196&44.08& $23.059\pm 0.118$& $21.544\pm 0.156$\\
3 &08.985&49.23& $22.299\pm 0.072$& $21.079\pm 0.092$\\
4 &09.226&55.04& $22.378\pm 0.094$& $21.094\pm 0.131$\\
5 &09.876&55.91& $19.667\pm 0.011$& $18.360\pm 0.012$\\
6 &09.751&45.68& $22.416\pm 0.079$& $21.288\pm 0.124$\\
7 &07.329&37.07& $22.435\pm 0.075$& $21.261\pm 0.084$\\
8 &08.366&55.18& $21.235\pm 0.025$& $19.939\pm 0.023$\\
9 &08.929&38.63& $22.803\pm 0.111$& $21.535\pm 0.186$\\
10&09.480&46.34& $21.790\pm 0.045$& $20.836\pm 0.050$\\
11&07.552&35.55& $21.824\pm 0.050$& $20.651\pm 0.067$\\
12&09.643&54.90& $23.322\pm 0.219$& $21.956\pm 0.296$\\
13&08.416&56.06& $22.021\pm 0.053$& $21.072\pm 0.084$\\
\tableline
\end{tabular}
\tablecomments{
The errors quoted are the formal fitting
errors. The zero point errors are 0.05\,mag in both J and K$_{\rm s}$.
}
\end{table}

\clearpage

\begin{table}[t]
\smallskip
\caption[]{
{\em Chandra} X-ray Point Sources on S3 and their (tentative)
counterparts \label{tab:cxosources}}
\smallskip
\def\pha{\phantom{23 00 }}
\def\phd{\phantom{+58 52 }}
\begin{tabular}{lllllll}
\tableline
source\tablenotemark{a}& $\alpha_{J2000}$ &$\delta_{J2000}$
&$\sigma_{\alpha,\delta}$\\
\tableline
X1 ($1\farcs9$, 310)\tablenotemark{b}&23 00 33.389& +58 52 47.12&$0\farcs09$\\ 
1425-14436845\tablenotemark{c}         &\pha33.396& \phd47.16& $0\farcs2$\\ 
DSS1 \tablenotemark{d}&\pha33.393& \phd47.23& $0\farcs2$\\
DSS2 \tablenotemark{d}&\pha33.380& \phd47.04& $0\farcs2$\\
\vspace{1mm}
P60 \tablenotemark{d} &\pha33.402& \phd46.96& $0\farcs03$\\
X2 ($1\farcs4$, 297)\tablenotemark{b}&23 00 43.360& +58 50 30.44&$0\farcs06$\\
\vspace{1mm}
P60 \tablenotemark{d} &\pha43.373& \phd30.11& $0\farcs09$\\
\vspace{1mm}
X3 ($1\farcs1$, 66)   &23 00 46.351& +58 53 08.76&$0\farcs11$\\
\vspace{1mm}
X4 ($1\farcs2$, 32)   &23 00 55.044& +58 54 59.05&$0\farcs25$\\
X5 ($3\farcs5$, 37)   &23 00 59.943& +58 59 11.03&$0\farcs72$\\
1425-14449773\tablenotemark{c}         &\pha59.798& \phd11.68& $0\farcs2$\\
DSS1 \tablenotemark{d}&\pha59.816& \phd11.51& $0\farcs2$\\
\vspace{1mm}
DSS2 \tablenotemark{d}&\pha59.815& \phd11.69& $0\farcs2$\\
X6 ($2\farcs3$, 380)  &23 01 00.118& +58 57 31.01&$0\farcs11$\\
1425-14449921\tablenotemark{c}         &\pha00.032& \phd31.12& $0\farcs2$\\
DSS1 \tablenotemark{d}&\pha00.038& \phd31.00& $0\farcs2$\\
DSS2 \tablenotemark{d}&\pha00.107& \phd31.20& $0\farcs2$\\
\vspace{1mm}
P60 \tablenotemark{d} &\pha00.105& \phd31.17& $0\farcs03$\\
X7\tablenotemark{e} ($0\farcs9$, 27) &23 01 07.175& +58 49 41.62&$0\farcs09$\\
\tableline
\end{tabular}
\smallskip
\tablenotetext{a}{For the X-ray sources we give in parentheses we give
the approximate image size and number of counts detected.}   
\tablenotetext{b}{X1 and X2 are RP1 and RP2 of \cite{rhop97}
respectively.}  
\tablenotetext{c}{USNO-A2.0 identifier.}
\tablenotetext{d}{Epochs are 1953.830 for DSS1 and USNO-A2.0, 1989.668
for DSS2 and 2000.885 for P60.}
\tablenotetext{e}{There is an extended source at this position on the edge of a
plate taken with LRIS in 1997.}
\tablecomments{
All sources except X5 were on the Palomar
images. Except for X1, X2 and X6, no object is found at the position
of any of these sources on these images. All X-ray positions were shifted by
$+0\farcs28$ in RA and $-0\farcs45$ in DEC to match the optical
positions.  The optical positions are on the system defined by the
USNO-A2.0 catalog.  Hence, they should be on the International
Celestial Reference System to $\sim\!0\farcs2$. Both the DSS1 and the
USNO-A2.0 positions are determined from the same plate. See text for
details.
}
\end{table}

\clearpage

\begin{figure*}
\caption{Cut outs of the I band image and K$_{\rm s}$ and J band mosaics of the
field around 1E~2259+58.6. A segment of the {\em Chandra} 99\% confidence
error circle is overplotted on the K$_{\rm s}$ band image, on all
other images the position is indicated by tickmarks. The best {\em
ROSAT} error circle of \citet{hullvkvk00} is overdrawn on the I band image for
reference. We note that although star~1 appears to consist of two dots
in K$_{\rm s}$, it is consistent with being a single point source.
\label{fig1}}    
\end{figure*}

\begin{figure}
\caption{
Broad band emission spectrum of 1E~2259+58.6. Denoted by CXO
are the X-ray data \citep{patekw+01}, plusses are absorbed 
and diamonds unabsorbed X-ray fluxes. Shown as dashed
lines are the blackbody and power-law components of the best fit
model. Also shown are the optical and infrared limits and the K$_{\rm
s}$ band detection, as observed and after correction for interstellar
reddening. \label{fig2}}   
\end{figure}

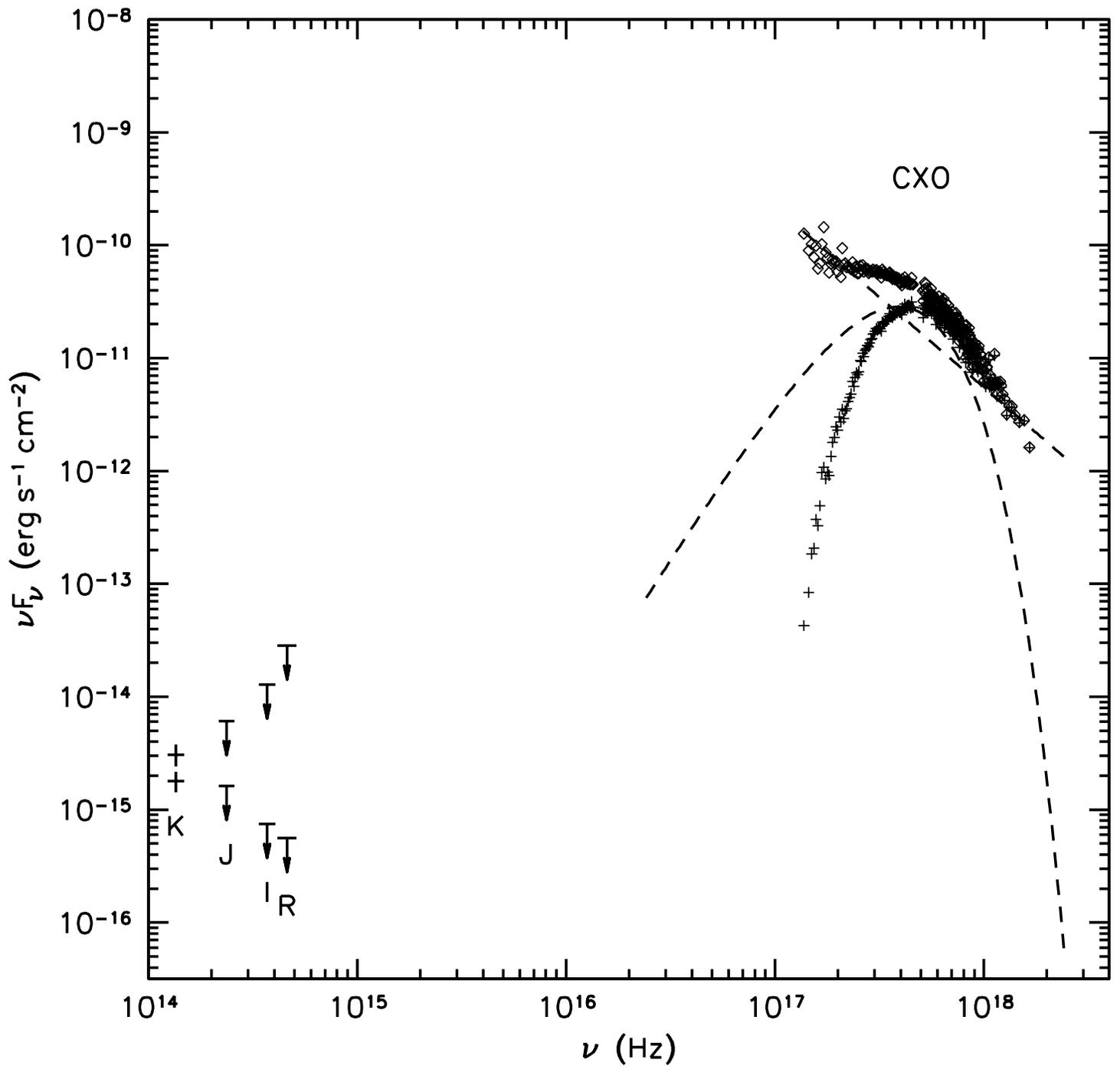
\end{document}